\documentstyle[aps,preprint]{revtex}
\newcommand{\be}{\begin{equation}}
\newcommand{\ee}{\end{equation}}
\newcommand{\Be}{\begin{eqnarray}}
\newcommand{\Ee}{\end{eqnarray}}
\newcommand{\f}{\frac}
\begin{document}
\tightenlines
%\draft
\title{Backreaction to wormhole by classical 
scalar field: \\
Will classical scalar field destroy wormhole?}
\author{
Sung-Won Kim\footnote{E-mail address: sungwon@mm.ewha.ac.kr 
}}

\address{Department of Science Education      
\\
Ewha Womans University
\\ 
Seoul 120-750, Korea}

\maketitle

\begin{abstract}

There are two effects of extra matter fields on 
the Lorentzian traversable wormhole.
The ``primary effect'' says that the extra matter can afford to be 
a part of source or whole source of the wormhole
when the wormhole is being formed. 
Thus the matter does not affect the stability of wormhole
and the wormhole is still safe.
If the extra matter is extotic,  
it can be the whole part of the source of the 
wormhole.

The ``auxiliary effect'' is that the extra matter  
plays the role of the additional 
matter to the stably-existed wormhole by the other exotic matter.
This additional matter will change the geometry of wormhole
enough to prevent from forming the wormhole by backreaction.
In the minimally coupled massless scalar field case,
the self-consistent solution was found.
The backreaction of the scalar field can dominate the exotic 
matter part so that it will hinder the formation of the wormhole.

\end{abstract}


\newpage

\section{Introduction}

One of the most important  issues in making 
a practically usable Lorentzian wormhole
is just the traversability\cite{MT88,MTY88}.
If it is traversable, there is a good usability, 
such as the short-cut in space\cite{MT88},
the time machine\cite{MTY88,KT91}, and 
the inspector of the interior of black hole\cite{FN93}.  
For the realization problem, 
it is believed that the wormhole is created in Planckian era
as the quantum foam\cite{V95}. 
As the other approach about wormhole
creation,  it is suggested that the wormhole can be made 
from a black hole by adding some special exotic matter\cite{H98}. 
The wormhole model as the final state of black hole evaporation was
very recently established in two dimensions\cite{KL99}.  

To make a Lorentzian wormhole traversable, one has usually  used an exotic 
matter which violates the well-known energy conditions\cite{MT88,MTY88}.
For instance, a wormhole in the inflating cosmological model still requires the 
exotic matter to be traversable and to maintain its shape\cite{R93}.
It is known that the vacuum energy of the inflating wormhole does not 
change the sign of the exoticity function.
A traversable wormhole in the Friedmann-Robertson-Walker(FRW) 
cosmological model, however, does not necessarily require the
exotic matter at the very early times\cite{K96}. The result means that there
were an exotic period in the early universe.

The problem about the maintaining wormhole by other fields 
relating with the exotic 
property has also been interesting to us.
There are two ways to generalize or modify the Lorentzian traversable
wormhole spacetime. 
(From now on `wormhole' will be simply
used as the meaning of the `Lorentzian traversable wormhole' 
unless there is a confusion.) 
One way is the generalization of the wormhole 
by alternative theory, for example,
Brans-Dicke theory, Einstein-Cartan theory, etc.
The other is the generalization by adding the extra matter.

In the case of the latter generalization, the added matter
will play two kinds of roles in affecting the wormhole spacetime.
The first role of  the 
other physical matters (for example,
scalar field, charge, spin, etc.)
 is the ``primary effect'', which says that the added matter is
the partial or total source of the wormhole.
The matter gets involved in the constructing stage of the wormhole.
The wormhole is safe under the addition of the
matter, since the matter is 
a part of the sources for constructing wormhole.  
This means that if
this added extra matter field has the exotic properties,
it shares the exoticity with other matter
which is exotic.
When the other fields are not exotic,
the added matter will monopoly the exotic property. 
In this case, wormhole
cannot be performed without this extra matter.
Its example is the case of the wormhole solution with 
scalar  field\cite{KK98}.

The second role is the ``auxiliary effect''. 
In this effect, the added matter is not a source,
but an extra effect to the existed wormhole.
Therefore, it is not involved with the constructing
stage of the wormhole, but relates with 
the wormhole after the construction. 
Since the matter makes the extra geometry, 
the wormhole is not safe when this effect 
dominate the exotic matter for constructing wormhole. 
It is the backreaction to the wormhole by the 
additional field.

In this paper, the self-consistent solution 
of the wormhole with classical, minimally-coupled, 
massless scalar field is found.  
The backreaction of the scalar 
field on wormhole spacetime is also found 
to see the stabilities of the wormhole.
The result is that the modified wormhole can be broken 
by the large variation of the scalar field.

The similar works about the scalar field effect 
on black hole and wormhole have been done by several
authors.
For the example on black hole,
Fonarev\cite{F95} generated new exact solutions
of Einstein-scalar field equations from static vacuum
solutions of Einstein equations.  In that paper,
the cosmological black hole solution was obtained.

It was already shown that the minimally coupled 
massless scalar field taken as
a source of Einstein gravity admitted
only the Schwarzschild black hole as a solution\cite{CB70}.
The result for the conformally invariant system was same\cite{Z95}.
Chamber et al\cite{CHT97} tried to find the evolution
of a Kerr black hole emitting purely
scalar radiation via Hawking process.

There were also some works for the effect on wormholes.
Taylor and Hiscock\cite{TH97} examined whether the
stress-energy of quantized fields in fact will have
the appropriate form to support a wormhole geometry. They
do not attempt to solve the self-consistent semiclassical
Einstein equations. They found that the stress-energy tensor of the
quantized scalar field is not even qualitatively of the
correct form to support the wormhole.
To maintain a wormhole classically, Vollick\cite{V97} found
the effect of the coupling a scalar field to matter which
satisfies the weak energy condition.

In Sec. II, these ``primary'' and ``auxiliary''
effects are described more concretely
in analytic form.
Breakdown of the wormhole by the backreaction of classical
scalar field is calculated in Sec. III.
Finally the summary and the further problems are discussed
in Sec. IV.

\section{Generalization of wormhole by extra matter}

The Einstein equation for the simplest normal
(usual) wormhole spacetime is given as
\be
G^{(0)}_{\mu\nu} = 8\pi T^{(0)}_{\mu\nu} 
\label{eq:ein}
\ee
The left hand side, $G^{(0)}_{\mu\nu}$, 
is the wormhole geometry
and the right hand side, $T^{(0)}_{\mu\nu}$,
is the exotic matter 
violating the known energy conditions, which is needed
to construct the wormhole.

The Einstein equation with the ``primary effect''  is
\be
G^{(0)}_{\mu\nu}=8\pi[T_{\mu\nu}+T^{(1)}_{\mu\nu}]
=8\pi T^{(0)}_{\mu\nu}
\label{eq:ein1} 
\ee
While the left hand side is usual wormhole geometry,
the right hand side is divided into two terms: the original matter
$T_{\mu\nu}$ and the added matter $T^{(1)}_{\mu\nu}$.
Since the sum of the right hand side 
is still $T^{(0)}_{\mu\nu}$, the added matter does not affect
on the geometry of the wormhole any more. 
The original matter
part, $T_{\mu\nu}=T^{(0)}_{\mu\nu}- T^{(1)}_{\mu\nu} $,
will be determined according to the magnitude of
the additional matter $T^{(1)}_{\mu\nu}$. 
When $T^{(1)}_{\mu\nu}$ is not exotic,
 the original matter part $T_{\mu\nu}$ should be exotic.
If the additional matter $T^{(1)}_{\mu\nu}$ is 
exotic, then $T_{\mu\nu}$ is not 
necessarily exotic.  
In this case, we can construct
wormhole with usual matter, 
while the exotic part
will be replaced by the additional $T^{(1)}_{\mu\nu}$, which makes 
total matter $T^{(0)}_{\mu\nu}$ exotic.
The special model of this case will be 
$G^{(0)}_{\mu\nu}=8\pi T^{(1)}_{\mu\nu}$, i.e., 
the additional matter takes over the whole $T^{(0)}_{\mu\nu}$ 
without any other matter. It was already
tried in scalar field case (if only $T^{(1)}_{\mu\nu}$ is
exotic), for example, the conformally-coupled case\cite{KK98}.

For the ``auxiliary effect'', the additional matter $T^{(1)}_{\mu\nu}$
is added 
to the right hand side of Eq.~(\ref{eq:ein}) and 
its backreaction $G^{(1)}_{\mu\nu}$
is added  to the geometry, 
the left hand side,
so that the Einstein equation becomes
\be
G^{(0)}_{\mu\nu}+G^{(1)}_{\mu\nu}=8\pi 
[T^{(0)}_{\mu\nu}+T^{(1)}_{\mu\nu} ].
\label{eq:ein3}
\ee
The sum of the matters of the right hand side already 
satisfies the conservation law.
However, there is no guarantee on the 
exoticity of the right hand side.
When the effect on the geometry of wormhole, $G^{(1)}_{\mu\nu}$,
is large enough to dominate any structure,
the additional matter, for example, scalar field
has the preventing mechanism from the formation 
of the wormhole.

There might be an interaction term $T^{\rm (int)}_{\mu\nu}$
between $T^{(0)}_{\mu\nu}$
and $T'_{\mu\nu}$ such as Ref.\cite{V97}.  
The model is the maintaining wormhole by the 
coupling a scalar field to matter that is not exotic,
but the coupling is not.
Then there also might be the interaction term 
$G^{\rm (int)}_{\mu\nu}$
in geometry.
In some case, $G^{\rm (int)}_{\mu\nu}$ may be joined
in Eq.~(\ref{eq:ein3}), 
even though $T^{\rm (int)}_{\mu\nu}$ does not appear,
so that the equation can have the form as 
$G^{(0)}_{\mu\nu}+G^{(1)}_{\mu\nu}+G^{\rm (int)}_{\mu\nu}
=8\pi (T^{(0)}_{\mu\nu}+T^{(1)}_{\mu\nu})$.
This example is the charged wormhole case which 
will be discussed in separate paper\cite{KL}.

\section{Backreaction to wormhole} 
       
\subsection{Static wormhole}

Before we study the wormhole model with scalar field,
the metric of the static wormhole is given by
\be 
ds^2 = -e^{2\Lambda(r)}dt^2 +
\f{dr^2}{1-b(r)/r}
+ r^2 (d\theta^2+\sin^2\theta d\phi^2). 
\label{eq:metric}
\ee
The arbitrary functions $\Lambda(r)$ and $b(r)$ are lapse and 
wormhole shape functions, respectively.  The shape of the wormhole
is determined by $b(r)$. Beside the
spherically symmetric and static spacetime,  
we further assume a zero-tidal-force as seen by
stationary observer, $\Lambda(r)=0$,  to make the problem simpler. 

The Einstein equation Eq.~(\ref{eq:ein}) is given as
\Be
\f{b'}{8\pi r^2} &=& \rho^{(0)}, 
\label{eq:worm11} \\
\f{b}{8\pi r^3} &=& \tau^{(0)}, 
\label{eq:worm12} \\
\f{b-b'r}{8\pi r^3} &=& P^{(0)}. 
\label{eq:worm13}
\Ee      
Assuming a spherically symmetric spacetime, one finds
the components of $T_{\hat{\mu}\hat{\nu}}^{(0)} $ in orthonormal 
coordinates
\be
T^{(0)}_{\hat{t}\hat{t}} = \rho^{(0)}(r),
~~T^{(0)}_{\hat{r}\hat{r}} = -\tau^{(0)}(r),
~~T^{(0)}_{\hat{\theta}\hat{\theta}} = P^{(0)}(r),
\label{eq:matter}
\ee
where $\rho^{(0)}(r), \tau^{(0)}(r)$ and $P^{(0)}(r)$ are  the 
mass energy density, radial tension per unit area, and lateral 
pressure, respectively, as measured by an observer at
a fixed $r, \theta, \phi$. 
 
\subsection{Primary effect by scalar field}

Firstly, we study the simplest case of a static Lorentzian wormhole with a 
minimally-coupled massless scalar field.
The additional matter Lagrangian due to the scalar field is
given by
\be
{\cal L} = \f{1}{2}\sqrt{-g}g^{\mu\nu}\varphi_{;\mu}\varphi_{;\nu} 
\label{eq:lag}
\ee
and the equation of motion for $\varphi$ by
\be
\Box\varphi=0. \label{eq:wave}
\ee
The stress-energy tensor for $\varphi$ is obtained from Eq. (\ref{eq:lag}) as
\be
T_{\mu\nu}^{(1)} = \varphi_{;\mu}\varphi_{;\nu}
-\f{1}{2}g_{\mu\nu}g^{\rho\sigma}\varphi_{;\rho}\varphi_{;\sigma}.
\label{eq:energy}
\ee
Now the Einstein equation Eq.~(\ref{eq:ein1}) has an additional 
stress-energy tensor
(\ref{eq:energy})
\be
G^{(0)}_{\mu\nu} = R_{\mu\nu}-\f{1}{2}g_{\mu\nu}R = 8\pi T^{(0)}_{\mu\nu}
= 8\pi(T_{\mu\nu}+T_{\mu\nu}^{(1)}),
\label{eq:ein5}
\ee
where $T_{\mu\nu} $ is the stress-energy tensor of the
background matter.
Here $T_{\mu\nu}^{(1)}$ will be a portion of $T_{\mu\nu}^{(0)}$ 
as the ``primary effect'' to the
wormhole.
One also finds
the components of $T_{\hat{\mu}\hat{\nu}}$ in orthonormal 
coordinates
\be
T_{\hat{t}\hat{t}} = \rho(r),
~~T_{\hat{r}\hat{r}} = - \tau(r),
~~T_{\hat{\theta}\hat{\theta}} = P(r),
\label{eq:matter1}
\ee
as the similar way of Eq.~(\ref{eq:matter}).
Thus not only the scalar field $\varphi$ but also the matter 
$T_{\mu\nu}$ are assumed to depend only on $r$.
The components of $T_{\mu\nu}^{(1)} $ in the static wormhole metric 
(\ref{eq:metric}) with $\Lambda = 0$ have the form
\Be
T_{tt}^{(1)}&=&\f{1}{2}\left(1-\f{b}{r}\right)\varphi'^2, 
\label{eq:scem1}\\
T_{rr}^{(1)}&=&\f{1}{2}\varphi'^2, \label{eq:scem2}\\
T_{\theta\theta}^{(1)}&=&-\f{1}{2}r^2\left(1-\f{b}{r}\right)
\varphi'^2, \label{eq:scem3}\\
T_{\phi\phi}^{(1)}&=&
-\f{1}{2}r^2\left(1-\f{b}{r}\right)\varphi'^2\sin^2\theta, 
\label{eq:scem4}
\Ee
where and hereafter a prime denoted the differentiation with
respect to $r$.  In the spacetime with  the metric (\ref{eq:metric}), 
the field equation Eq.~(\ref{eq:wave}) of $\varphi$ becomes
\be
\f{\varphi''}{\varphi'}+\f{1}{2}\f{(1-b/r)'}{(1-b/r)}+\f{2}{r} = 0 \quad
\quad \mbox{or} \quad \quad
r^4\varphi'^2\left(1-\f{b}{r}\right) = \mbox{const} ,
\label{eq:phi}
\ee
and the Einstein equations are given explicitly by
\Be
\f{b'}{8\pi r^2}&=& \rho^{(0)}=\rho + \f{1}{2}\varphi'^2\left(1-\f{b}{r}\right), 
\label{eq:scein1}\\
\f{b}{8\pi r^3}&=& \tau^{(0)}=\tau - \f{1}{2}\varphi'^2\left(1-\f{b}{r}\right), 
\label{eq:scein2}\\
\f{b-b'r}{16\pi r^3}  &=& P^{(0)} = P - \f{1}{2}\varphi'^2\left(1-\f{b}{r}\right).
\label{eq:scein3}
\Ee
Thus one sees that  the conservation law of the 
effective  
stress-energy
tensor $T_{\mu\nu}+
T^{(1)}_{\mu\nu}$ still obeys the same equation
\be
\tau'^{(0)}+\f{2}{r}(\tau^{(0)}+P^{(0)}) = 0.
\ee
With the equation of state, $P^{(0)}=\beta\rho^{(0)}$ and
the appropriate asymptotic flatness, 
we find the matter as functions of $r$ \cite{K96,KK98}
\Be
\rho^{(0)}(r) &\propto& r^{-2(1+3\beta)/(1+2\beta)}, \label{eq:sol1}\\
\tau^{(0)}(r) &\propto& r^{-2(1+3\beta)/(1+2\beta)}, \label{eq:sol2}\\
b(r) &\propto& r^{1/(1+2\beta)} , \qquad \beta < -\f{1}{2}.
\label{eq:sol3}
\Ee
Here the wormhole is stable, because the added scalar field
does not affect any spacetime geometry.

\subsection{Auxiliary effect by scalar field}

Next we think the ``auxiliary effect'' by the scalar fields,
the additions of the scalar field $T^{(1)}_{\mu\nu}$
to the existing wormhole matter $T^{(0)}_{\mu\nu}$. 
If we add $G^{(1)}_{\mu\nu}$ as
an additional geometry to $G^{(0)}_{\mu\nu}$ by scalar field, the Einstein 
equation Eq.~(\ref{eq:ein3}) is changed 
from Eq.~(\ref{eq:worm11})-(\ref{eq:worm13}) into
\Be
\f{b'}{8\pi r^2} + \f{1}{8\pi}\f{\alpha}{r^4} &=& \rho^{(0)} + \f{1}{2}\varphi'^2
\left( 1 - \f{b}{r} \right), 
\label{eq:worm21} \\
\f{b}{8\pi r^3} - \f{1}{8\pi}\f{\alpha}{r^4} &=& \tau^{(0)} - \f{1}{2}\varphi'^2
\left( 1 - \f{b}{r} \right),
\label{eq:worm22} \\
\f{b-b'r}{8\pi r^3} - \f{1}{8\pi}\f{\alpha}{r^4} &=& P^{(0)} - \f{1}{2}\varphi'^2
\left( 1 - \f{b}{r} \right),
\label{eq:worm23}
\Ee      
when the interaction between the matter and additional
scalar fields is neglected.
Here $\alpha$ is defined as the positive value.
The term  $\alpha/r^4$ is added to the left hand side, because  
the field equation Eq.~(\ref{eq:phi}) of field $\varphi$
shows that  $\varphi'^2 \left( 1 - \f{b}{r} \right) \propto r^{-4}$. 

If we set $b_{\rm eff} = b - \alpha/r$ instead of $b$, then
the Einstein equations Eq.~(\ref{eq:worm21})-(\ref{eq:worm23}) satisfy
self-consistently.
Thus the effect by the scalar field on the wormhole is
only by changing the wormhole function
$b$ into $(b-\alpha/r)$ without any interaction term in the 
left hand side. 
Since $b$ is proportional to $r^{1/(1+2\beta)}$,
with the proper parameter $\beta$ of equation of state,
the wormhole will vary seriously by the additional factor 
$-\alpha/r$, according to the values of the parameters 
$\beta$ and $\alpha$. 
While $\beta$ is given as the equation of state by the
choice of the appropriate matter, 
$\alpha$ depends on the changing 
rate of the scalar field $\varphi$, if only
$b$ is fixed.

For the wormhole function, we can set as
\be
b = b_0^{\f{2\beta}{2\beta+1}}r^{\f{1}{2\beta+1}}, 
\ee
where the value of $\beta$ should be less than $-\f{1}{2}$
so that the exponents of $b_0$ and $r$ can be negative.

When $-1<\beta<-\f{1}{2}$, the power  of $b(r)$ is 
less than $-1$, the power of the second term,
so $b(r)$ vanishes more quickly in the far region. 
Thus it gives the negative region for $b_{\rm eff}$ at large $r$, 
even though it has the positive regions near throat when
$b_0>\alpha$. 
The region of the positive $b_{\rm eff}$ is $b_0<r<r_0$, where
$r_0 = \alpha^{(1+2\beta)/(2+2\beta)}/b_0^{\beta/(1+2\beta)}$.
If $b_0 < \alpha$, $b_{\rm eff}$ is negative at all $r$, 
which is not suitable for wormhole formation.

If only $\beta \le -1$ and $b_0 > \alpha$, the wormhole
is safe, because $b_{\rm eff}$ is positive at all $r$. 
Otherwise, the scalar field effect will change
the wormhole structure into others,
since the scalar field dominates the 
exotic matter.
Thus the addition of the minimally-coupled, massless
scalar field does not guarantee the structure
of wormhole.

Now we shall examine the special case of this backreaction problem,
for instance, $\beta = -1$ which is $b = b_0^2 /r$.
In this case, the solution of the scalar field  is given as\cite{KK98}
\be
\varphi = \varphi_0 \left[ 1 - \cos^{-1} \left( \f{b_0}{r}
\right) \right].
\ee
Thus the proportional constant $\alpha$ becomes
\be
\alpha = \varphi_0^2 b_0^2 ,
\ee
where $b_0$ is the minimum size of the wormhole and 
$\varphi_0$ is the value of $\varphi(r)$ at $r=b_0$.
Therefore, $\varphi_0 < b_0^{-1/2}$ is the condition that 
is required for maintaining the wormhole under the
addition of the scalar field. 
In this choice of $\beta = -1$, there is no $r_0$ at
which $b_{\rm eff}$ changes from positive value to negative
one.

We can also apply the result to the other form
of $b(r)$, which means the exotic matter distribution  
in the restricted region only, 
``absurdly benign'' wormhole,
\be
b(r) = \left\{ \begin{array}{cl}
b_0 [1-(r-b_0)/a_0]^2, \Phi(r) = 0, & ~~{\rm for}~~~~b_0 
\le r \le b_0 + a_0, \\
b=\Phi=0, & ~~{\rm for}~~~~ r \ge b_0+a_0
\end{array}
\right.
\ee
In this case, since the second term $-\alpha/r$ in the 
effective shape function by the scalar 
field extends to over the region $r \ge b_0 + a_0$,
there will be a negative $b_{\rm eff}$ within this range of values of
$r$, which means that wormhole will be broken.

\section{Discussion}

Here we studied the backreaction to wormhole by the
scalar field and found the self-consistent solution. 
The scalar field effect may break the wormhole structure
when the field and the variation of the field is large.
Similar consequences are obtained in charged wormhole case\cite{KL},
in which there is the interaction term in geometry, even though
no interaction term in matter.                                     
It is natural that the addition of the nonexotic matter will
break wormhole if the ``auxiliary effect'' is large.

In this paper, the interaction between the extra field
and the original matter is neglected. 
If the interaction exists and it is large, it can change the whole 
geometry drastically.
If it is very small, it does not change the main structure of the wormhole.
The detailed discussion on these interactions will be in separate paper.

\acknowledgements

This research was supported by the 
MOST through National Research Program(98-N6-01-01-A-06)
for Women's University.

\end{document}